# Between Flat-Earthers and Fitness Coaches: Who is Citing Scientific Publications in YouTube Video Descriptions?

Olga Zagovora* (RPTU Kaiserslautern-Landau, Landau; DFKI GmbH, Kaiserslautern),
Katrin Weller (GESIS – Leibniz Institute for the Social Sciences)

* corresponding author: olga.zagovora@rptu.de

## Abstract

In this study, we undertake an extensive analysis of YouTube channels that reference research publications in their video descriptions, offering a unique insight into the intersection of digital media and academia. Our investigation focuses on three principal aspects: the background of YouTube channel owners, their thematic focus, and the nature of their operational dynamics, specifically addressing whether they work individually or in groups.

Our results highlight a strong emphasis on content related to science and engineering, as well as health, particularly in channels managed by individual researchers and academic institutions. However, there is a notable variation in the popularity of these channels, with professional YouTubers and commercial media entities often outperforming in terms of viewer engagement metrics like likes, comments, and views. This underscores the challenge academic channels face in attracting a wider audience.

Further, we explore the role of academic actors on YouTube, scrutinizing their impact in disseminating research and the types of publications they reference. Despite a general inclination towards professional academic topics, these channels displayed a varied effectiveness in spotlighting highly cited research. Often, they referenced a wide array of publications, indicating a diverse but not necessarily impact-focused approach to content selection.

**Keywords**: **altmetrics, social media metrics, YouTube, academic videos**

## 1. Introduction

In this work, we take a close look at how videos on YouTube are connected to scholarly publications via references in their video description text. We are specifically studying who is using links to scholarly publications in their video descriptions, and in this way want to learn more about the role of the general YouTube community in discourses about research findings. Our work is embedded in the broader context of studying new forms of engaging with scholarly communications in different online environments which is also referred to as altmetrics (Haustein, 2016; Priem et al., 2010; Sugimoto et al., 2017). Altmetrics as an area of research is often dedicated to identifying and exploring new ways of measuring the impact of scientific work - outside of traditional scholarly communication structures. Those traditional structures are mainly focused on formal publications (journal articles, books, proceedings etc.) and the citations that they receive from other published work, which are interpreted as indications of information flows between the different publications (for example described by Cronin (1984)). Altmetrics data in contrast may be looking into additional sources outside the academic publishing environments to identify how a community beyond academia may be referring to scientific work. Often altmetrics data is collected from social media platforms that support networking and communication across a potentially world-wide user community (such as Twitter (Haustein, 2019)) or other community driven online platforms that enable collaborative content generation (such as Wikipedia (Kousha & Thelwall, 2017)). User communities of these types of platforms are considered as representatives of a general public audience beyond academia, and platform data about scholarly work being mentioned, commented on or otherwise used in interactions is used to measure scholarly impact within this given platform. Other altmetrics sources may include bookmarking platforms or news outlets, as additional ways in which academic research can be referenced outside of traditional citation structures. Overall, altmetrics is not only a research field that aims to learn more about how research results are perceived beyond academia, the term is also used for a line of practical applications and tools that assign new types of indicators to rate publications' performance, typically based on the quantity of mentions of a publication. Altmetrics data is collected (and sold) by so-called aggregators, the most prominent currently being Altmetric.com[1], PlumX[2], and CrossRef Event Data[3]. The indicators they provide are applied in different settings such as publishers' sites or repositories, where they may be used to advertise "impactful" publications. The role of these tools and indicators in recommending and measuring scholarly work makes it even more important on the other hand to better understand what kind of activities are being traced and measured. We need to learn more about whose activities are counted as altmetrics data, and eventually how (or even why) different user groups engage with scholarly works in public online spaces.

Within this context, YouTube is still rather understudied as an area for engaging with scholarly output and as a potential source for altmetrics data – despite the overall importance of YouTube as one of the most used Web platforms. Moreover, according to Mehrazar et al. (2018) around 20-30% of researchers actively utilize YouTube in their daily work. Some aspects that have been studied include TED Talks on YouTube, which represent a stellar example of how digital platforms can be utilized to distribute presentations related to science and technology directly from event venues (Sugimoto et al., 2013). Despite the fact that academics only constituted 21% of the speakers (Sugimoto et al., 2013), videos featuring academic presenters garnered more positive engagements, as evidenced by the higher number of 'likes' they received compared to videos with non-academic presenters (Sugimoto & Thelwall, 2013). While according to Sugimoto et al. (2013) delivering a TED Talk doesn't necessarily correlate with an increase in citation count for an individual researcher, data shows that speakers from academia tend to have a citation impact above the field mean. This underscores the potential of utilizing videos and their associated popularity metrics as a measure of academic achievement. At the same time, Thelwall et al. (2012) have advocated for the use of view counts adjusted for audience size as a valuable metric for assessing video impact. Additionally, YouTube comments and videos have proven to be rich sources for various research purposes, including evaluation of gender bias within channels that are related to science (Thelwall & Mas-Bleda, 2018), topic mining (Thelwall, 2017), and understanding audience responses to significant issues (Thelwall, Sud, et al., 2012). The videos that were cited in published academic research have also been the subject of analysis by Kousha et al. (2012).

---

[1] https://www.altmetric.com/

[2] https://plumanalytics.com/

[3] https://www.crossref.org/services/event-data/

The prior study (Zagovora & Weller, 2018) established a correlation between video view counts and citation impact, hinting at the potential of using these view counts as altmetrics to indicate research impact. This finding is further corroborated by Shaikh et al. (2023), who emphasize the importance of video view counts as a key predictor of citation frequency for research articles. Such trends not only elevate the popularity of these articles but also significantly enhance engagement with scientific content among the public. Moreover, Shaikh et al. (2023) propose that research articles which receive frequent mentions in tweets and substantial coverage in the news are more likely to be cited in videos, highlighting the interconnected nature of digital media and scholarly impact.

Numerous studies, including those by Zhang et al. (2023), Haustein (2019), and Tsou et al. (2015), have consistently shown that the bulk of altmetric data primarily originates from academic contributors and other scholarly entities, including research organizations, funding bodies, and academic journals. This trend underscores the significant role of the scholarly community in disseminating scientific content also in online platform environments, rather than a general public engagement with scientific content. Tsou et al. (2015) emphasize this point, indicating that the majority of tweets linking to scientific papers are predominantly authored by members of the academic community. This finding suggests a concentrated effort within the scholarly sphere to promote and share research findings. Further elaborating on this, Zhang et al. (2023) revealed a nuanced landscape where, although about half of the individuals tweeting about scientific papers are not directly affiliated with academic institutions, the original tweets about these papers mostly originate from within the academic community. Moreover, a significant portion of tweets from non-academic sources are often in response to these academic-originated tweets. Interestingly, it's also observed that only about half of the publications discussed on Twitter are mentioned by those outside the academic sphere (Zhang et al., 2023).

This pattern of engagement, where academic and non-academic contributions to altmetrics are intertwined yet distinct, offers valuable insights into the dynamics of research dissemination. It highlights the pivotal role of academia in initiating discussions about scientific work, which are then further propagated or responded to by a broader audience. Understanding this dynamic can enhance how altmetrics are interpreted and utilized, emphasizing the need to consider the source of altmetric data when evaluating the reach and impact of scholarly work.

However, there is a notable gap in these studies, as they primarily focus on platforms like Twitter and Mendeley. The role of YouTube, especially in terms of how actors engaged with research publications contribute to altmetrics, is still not thoroughly studied. This gap presents an opportunity for future research to explore how YouTube, as a platform with a distinct user base and content style, contributes to the dissemination and public engagement with scholarly work. Understanding YouTube's unique role in the altmetrics landscape could offer deeper insights into the multi-dimensional nature of how research is communicated and perceived in the digital age, thereby enriching the overall understanding and application of altmetrics in assessing the reach and impact of scholarly work in both academic and public spheres. Shaikh et al. (2023) have contributed a notable step towards better understanding the citation practices on YouTube. With this data, we extend the knowledge about citations on YouTube by adding a specific focus on the actors who link videos to scientific topics.

With our work we want to take a closer look behind the scenes of connections between YouTube videos and scholarly publications, trying to uncover general patterns as well as specific insights into topical foci, actors who make use of references to scholarly publications, and reactions by a broader audience. We do so by using a dataset of YouTube videos that each include at least one link to at least one scholarly publication (such as journal articles or books) in their video description. We assume that this kind of link resembles a form of "citation" between videos and publications, i.e. the video creators are listing a specific publication as a reference for their content. We have to note, however, that one limitation of this work is that the connection is specifically between the video *description text* and the publication, we did not use the actual video content material (e.g., via video transcripts), which might include additional layers of talking about scientific research and linking to scholarly publications. YouTube videos featuring citations of at least one scholarly paper in their descriptions were compiled using dataprovider Altmetric.com[4]. This provider offers information on different online platforms and the scholarly publications they are connected to. Another limitation of this work therefore is the dependence on Altmetric.com as the provider of the data, and the insights they share about how exactly they identify a resource being a scholarly publication. According to them this includes monitoring specific academic publishing domains and tracing persistent document identifiers (such as DOIs). Details on the collection procedure are provided in section 2.

Our data enables us to combine quantitative descriptions of the big picture, as well as a closer look at selected samples, therefore arriving at two objectives/dimensions for this paper:

**RQ1:** Which YouTube channels reference scholarly publications in their video descriptions?
  **RQ1.1** What are the topics of channels that reference publications? And how does this relate to video topics and publication topics?
  **RQ1.2** Is a general public (beyond people in academia) involved in sharing links to scholarly publications on YouTube?
  **RQ1.3** Are these channels a product of group or individual efforts?
**RQ2**: Do academic actors share different content than other YouTubers?

In Section 1 we provide a general descriptive view on our main dataset. Overall, 11,332 distinct YouTube channels that posted about 35,049 unique videos are represented in our data (after removing 6,396 videos that that were no more available on YouTube since they were deleted or banned at the point of data gathering). The distribution of videos per channel is skewed: 8,189 channels uploaded just a single video citing scholarly publications, while the two most "productive" channels in our data accounted for 1,134 ("ScienceVio") and 681 ("Seeker") videos that reference at least one publication. We offer additional insights on the full dataset in sections 2 and 5, and use smaller subsets for manual coding of channel types.

---

[4] https://www.altmetric.com/

To answer our RQs we developed a coding scheme for manually annotating selected channels in different dimensions: a) channel types are sorted into different categories including, amongst others, categories for accounts by professional academics, regular YouTube contributors with or without identifiable academic degrees, media, and organizational accounts; b) channel content was classified into topical areas such as, amongst others, health, fitness, different scientific fields (professional science content and popular science videos), creative content, politics, opinions. In a group of three annotators, accounts were coded according to these categories for the following samples from the main dataset (see details in section 2): YouTube channels associated with the top 100 most viewed, 100 most liked, 100 most commented, with the most emotionally discussed videos, as well as 100 random channels. We were also able to sample the 100 most disliked videos from the dataset, since data collection happened before YouTube discontinued the dislike feature[5].

Channel types include recurring themes, for example, content creators that focus on producing series of videos that contain easily understandable explanations for scientific phenomena from various areas, channels that offer personal opinions and world views, and channels from different producers of online fitness coaching that back up their content with links to scientific findings. Other cases are more unique in their appearance, such as individuals privately engaging with scientific publications as part of their videos. We also encountered videos as official supplementary material for the cited papers. The full procedure and results will be presented in sections 3 and 5.

# 2. Data collection setup and initial dataset description

For collecting the data for this project we used data access via Altmetric.com. Altmetric.com[6] is one of the altmetrics aggregation tools that provide mentions of research outputs from different sources[7] (i.e., social media, mainstream media, etc). All data was collected as of December 2017. At this point, we received all YouTube videos that included references to publications in their textual video description (i.e. not in the actual video). Here, Altmetric.com resolves it via direct links (i.e., URLs) to a research output that is placed in the description section. Altmetric.com maintains its own matching between research output and its metadata (Crossref-registered DOI, ISSN, PDF URLs, etc)[8] and resolves disambiguation. In this way, we received a list of 41,445 distinct YouTube videos posted in the years 2006-2017 that included one or more references in their video descriptions, with a total of 83,924 links to scholarly outputs. Across various video descriptions, certain academic publications were cited multiple times, amounting to a total of 51,743 academic publications being referenced. This total includes 48,527 articles, 2,901 books, and 315 book chapters.

---

[5] They were suspended as of December 13th, 2021. Details can be found here
https://support.google.com/youtube/thread/134791097/update-to-youtube-dislike-counts?hl=en
[6] https://www.altmetric.com/about-us/our-data/how-does-it-work/
[7] https://www.altmetric.com/about-us/our-data/our-sources/
[8] Find more details here: https://help.altmetric.com/support/solutions/articles/6000240585-scholarly-identifiers-supported-by-altmetric, https://help.altmetric.com/support/solutions/articles/6000240582-required-metadata-for-content-tracking,
https://help.altmetric.com/support/solutions/articles/6000240591-details-pages-creation-and-crossref-metadata,

For the YouTube videos in our collection, additional metadata about the videos, channels, and comments to those videos were obtained from Google's YouTube Data API v3[9]. We were thus able to consider video features such as video category as well as view, comment, like and dislike counts, and texts from comments on the videos.

By the time of data analysis, 6,396 videos were no longer accessible via the YouTube API. Those videos might have either been deleted by the platform (e.g., for the platform's rules violation) or removed by the channels that had posted them. Thus, we ended up with 35,049 distinct videos, published by 11.332 distinct YouTube channels. According to our analysis, 10 channels posted more than 500 videos (the maximum was 1,134 videos) that cite at least one academic publication (Figure 1), and 8,189 channels (i.e., 72%) uploaded just one video of this type (see Figure 1).

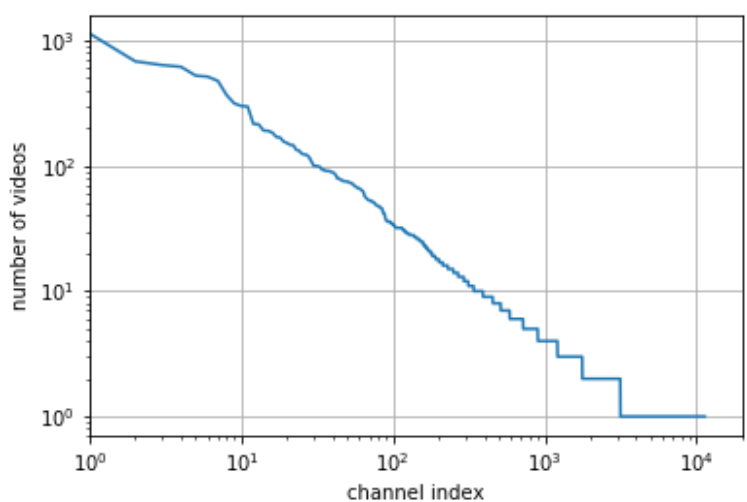

**Figure 1 - Distribution of research-related videos per YouTube channel (log-log plot)**

Thus, we were able to get data about 35,049 videos. Overall, 4,333,304 comments have been left for these videos by the time of data collection. All the video features (comments, dislikes, likes, views) follow the power law distribution (Figure 2), meaning that a few top videos received tremendous amounts of attention (e.g., one video from the dataset got 30 million views) and thousands of videos have not received any comments or likes. Only 15,613 videos received at least one comment, for 1899 videos the commenting option was disabled by the video channel that posted it.

We also supplemented each English comment with the polarity score (obtained using Python library TextBlob[10]). We had to reduce sentiment analysis to English comments (3,805,983 comments; 87,8% of comments) due to the lack of accuracy of polarity scores of the pre-trained model in non-English comments. In the dataset, 22.1% of comments were negative (polarity score from -1 to 0), 34.2% were neutral (polarity score equal to 0) and 43.7% were positive.

---

[9] https://developers.google.com/youtube/v3/docs?hl=en
[10] https://github.com/sloria/TextBlob

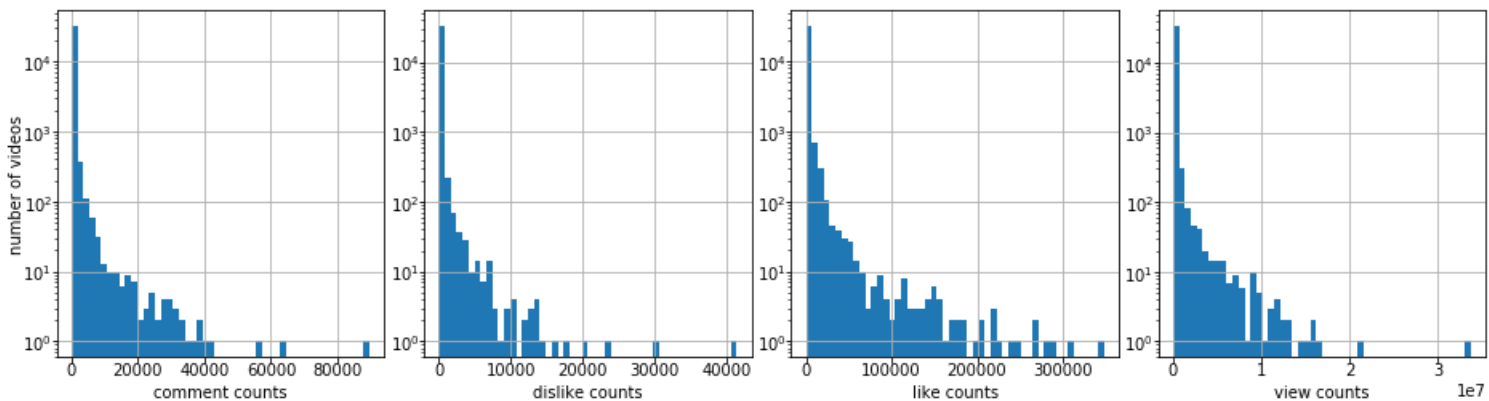

**Figure 2 - Counts for comments, likes, dislikes and views**

Moreover, citation counts of those publications that were referenced in one of the video descriptions were obtained using CrossRef API[11]. These values were normalized (based on publication year and Scopus subject) and log-transformed (Thelwall & Fairclough, 2017) to mitigate skewness and adjust for the accumulation of citations over time. Further in the paper, we will refer to these values as the *citation impact* of a publication.

Different subsets of YouTube channels have been created for closer investigation and as basis for the manual annotation procedure. We have considered some popularity metrics of videos (e.g., the number of comments to a video) and created the subsets of channels belonging to these videos. So, the following sets of channels were created:

a) a random set of 100 channels from our dataset - *random*,
b) 100 channels behind the top most commented videos - *top_commented*,
c) 100 channels behind the top most disliked videos - *top_disliked*,
d) 100 channels behind the top most liked videos - *top_liked*,
e) 100 channels behind the top most viewed videos - *top_viewed*,
f) 100 channels that referenced most cited publications - *top_citedpapers*,
g) 70 channels where the most negative discussions in the comments happened - *negative_discussion*,
h) 70 channels where the most positive discussions in the comments happened - *positive_discussion*.

The first set a) consists of 100 random channels sampled from the full list of unique channels in our dataset. The sets b)-e) were based on the popularity metrics of the video. It should be noted that in each of the categories, the aim was to generate a list of 100 channels; this does not correspond to the channels of the top 100 videos in each category, since often the same channels would have several videos in the top commented/liked/viewed categories. The set f) is based on the information about research publications. Here, for each publication that is cited by a video description, we calculated a log-transformed citation score (the score is normalized based on publication year and field of science and then log-transformed), and then all the publications were sorted based on those scores. Then the first 100 unique channels that posted these videos, where the highly cited researched publications were mentioned in the description, were selected.

---

[11] https://www.crossref.org/documentation/retrieve-metadata/rest-api/

To create the sets g) and h) the following procedure was in place: 1) For every comment in the dataset we calculated the sentiment score. 2) Then for every video with at least 10 comments, the average sentiment score of all comments was calculated. 3) We sorted these videos by the average sentiment scores and selected 70 top and bottom unique channels that posted these videos. For sets g) and f) we were not able to collect 100 channels because the average sentiment score was approaching 0, meaning that these videos have neutral comments on average, rather than negative or positive ones.

Since some popularity metrics of videos are co-correlated (e.g., videos with a high number of views would also likely have a high number of likes), we would get some common channels in some of the sets presented above. In order to estimate how many shared channels we have in each pair of sets, we calculated the Jaccard index (Table 1).

Jaccard similarity coefficient, in this case, shows what the percentage of shared channels is in each YouTube subgroup. So, top_viewed, top_liked, top_disliked, and top_commented groups share between 37% and 49% of videos.

**Table 1. Jaccard index** (or Jaccard similarity coefficient)

| | random | top_commented | top_disliked | top_liked | top_viewed | top_citedpapers | negative_discussion | positive_discussion |
|---|---|---|---|---|---|---|---|---|
| random | | 1.52 | 2.04 | 1.52 | 3.09 | 1.01 | 1.19 | 0 |
| top_commented | | | 41.84 | 40.85 | 36.99 | 5.26 | 4.94 | 0 |
| top_disliked | | | | 40.85 | 38.89 | 6.95 | 4.94 | 1.19 |
| top_liked | | | | | 49.25 | 6.38 | 4.29 | 0.59 |
| top_viewed | | | | | | 5.82 | 4.94 | 0.59 |
| top_citedpapers | | | | | | | 2.41 | 0 |
| negative_discussion | | | | | | | | 1.45 |
| positive_discussion | | | | | | | | |

# 3. Approach for Manually Coding Channel Types and Contents

To gain insights into RQs and to learn more about the channels posting videos with references to scientific publications we developed a coding scheme for manual coding of YouTube channels in our dataset. The coding scheme consists of two dimensions: a) the channel type and b) the channel topic(s). The authors of this paper created and continuously refined the coding scheme while reviewing the initial sets of channels from our data collection. This process involved conducting preliminary annotations on subsets of the data, beginning with a randomly selected subset. The final version of our coding scheme can be found as a shortened version in Table 2 and 3, listing the codes and short definitions. Three annotators (the authors of this paper and a student assistant) were coding the YouTube channels included in the subsets described above in section 2 (sets a-h), the exact coding procedure is described below.

With the **channel type** category, we aim at identifying what kind of actors are behind the YouTube channels in our dataset. Our research is mainly focusing on distinguishing actors with some relation to academia from a general YouTube community. This addresses the core foundation of altmetrics, where a central question is whether new types of data sources actually help to broaden the scope of measuring impact by including additional types of actors (beyond academics) who are interacting with scientific publications. In our context, one interesting aspect is to see if users from a broader, non-academic context are also using YouTube as a platform to engage with scholarly publications and scientific topics. On the other hand our coding scheme also includes more fine-grained categories covering different types of academia-related channels, including for example (groups of and individual) researchers, students, academic institutions or academic publishers. Channel categories remain rather broad on the level of the general public, where many channels are categorized as "YouTuber", i.e. users with a well established YouTube profile. For these cases, it is the combination with the topical categories that enables us to identify which areas of general public audiences are interacting with scientific publications, for example YouTube accounts in the context of gaming, or fitness and beauty topics.

The three coders independently coded the YouTube channels in the datasets a-h. Where necessary, we used translation tools to automatically translate texts (profile texts, video subtitles etc.) from languages that we were not familiar with and then worked with the English translations. To identify channel types based on actors involved in any given channel, the following procedure was applied:

> Step 1: The starting point was reading the channel description of a channel on YouTube and potentially watching some introductory videos on the channel if available.
> Step 2: We followed outlinks from the YouTube profile to additional profiles mentioned there (e.g., on other social media platforms such as Twitter or to personal or professional Websites) to also consider the information included therein.
> Step 3: We conducted a google search for the channel name or names of individuals mentioned in the channel description. This was mainly done to confirm academic affiliations on university websites and to check for additional external mentions of the channel, for example in news articles or blog posts.

With this procedure, the exact information found by each of the three annotators might have differed substantially, depending for example on which introductory YouTube videos were watched, what additional resources were found or how well one could make sense of different languages or their translations. In this situation, inter-coder agreement was naturally low and not sufficient to establish a useful category assignment, based on, e.g., majority votes. We therefore introduced another step in the procedure to reach a consensus for each channel in our annotation dataset:

> Step 4: For all channels in our sample sets we checked whether the tree annotators agreed on the channel type and for every channel where this was not the case, we discussed jointly until we arrived at a joint decision. Final decisions were typically influenced by pieces of evidence that one of the annotators found which others had not seen (e.g., a connection to an academic affiliation).

**Table 2. Categorization of YouTube channels** in our dataset based on their hosts.

| Abbreviation | Channel category | Description and examples |
|---|---|---|
| YOUTUBE | Active YouTubers | This category applies to channels managed by one or more individuals who produce YouTube videos either (semi-)professionally or on a regular basis and have an established online presence on the platform. This category includes channels that either have a channel description text that explains about the focus or the motivation behind the channel or have identifiable information from additional data sources (Wikipedia articles, links to social media and private web sites) about the people behind the channel. Moreover, the content should be "self-produced" videos, i.e. not just uploads of TV material, compilations. |
| YOUTUBEAC | YouTubers with academic background | This category represents a more detailed or precise variation of the above-mentioned YOUTUBER. It is intended for situations where one or more individuals who own the channel have been recognized as possessing an academic degree., However, these individuals are not necessarily affiliated with academic institutions as researchers. This is inclusive of all possible university degrees starting from bachelors degrees. This group also includes former researchers and former professional academics that no longer work in academia. |
| ACADEM | Professional academics | This category is used for Individuals or groups that can be recognized as professional academics, i.e. researchers (incl. PhD candidates) affiliated with an academic institution (such as universities, research institutes, also including industry research departments). For group accounts all members should have an academic affiliation. The category also includes teachers and lecturers affiliated with an academic institution. YouTube accounts from research teams, research projects and small research groups will be included in this category too. |
| STUDENT | Students | This category is used for cases where individuals behind the channel/videos identify as undergraduate or graduate students (or for cases where no individuals are present, but the videos display that they are part of a class assignment). |
| INSTITUTE | Academic institution | The YouTube channel is hosted by an institution (or its subdivision, e.g., department), such as universities, government agencies focusing on research, other research institutes, academic libraries. It also covers channels by (professional) academic associations (unless the focus of these associations and/or their YouTube presences is clearly on publishing, in those cases they will be categorized as publishers, see below category PUBL). The category also does not include research teams, projects and small research groups, as those will be included in the ACADEM category. |
| PUBL | Academic publisher | This category is designated for channels associated with academic publishers (such as Nature), preprint publishers, or academic journals. It also includes academic associations if they clearly act as publishers and/or their YouTube presence is clearly linked to their role in publishing members' works. |

| Abbreviation | Channel category | Description and examples |
|---|---|---|
| COMPANY | Commercial company or non-profit organization | This category is used for YouTube channels hosted by a commercial company (typically used to promote their service or products), or by other forms of organizations, including semi-commercial and non-profit. |
| MEDIA | Media / news publishers | This category is applied to channels run by commercial media companies or journalistic entities that have a well-established media presence beyond YouTube, but also operate a channel on the platform. This can be mainstream media, public broadcasting (including tabloid press), born-digital news outlets (such as buzzfeedvideo). It may also be individual journalists that affiliate with these types of media sources. |
| OTHER | Other | This category encompasses channels belonging to someone that do not align with the previously mentioned categories. |
| N.A. | no information available | It is applied in situations if there was no information accessible regarding the channel, such as when all channel's descriptions were not filled in, or when the persons behind the channel did not reveal their identity with at least either some name or their personal appearance in one of the videos. |

For each channel we also coded whether it was maintained by a **group of people or one individual person,** or unknown. Here the code (group, individual, unclear) was assigned based on the majority vote of the three coders.

With **channel topic(s)** categories, we further describe the core topical areas of the videos featured in a channel. A maximum of two topics could be assigned per channel, thus capturing the fact that several channels would include videos with different topical foci. Again we iteratively created the categorization scheme while looking into sets of random channels. Categories were refined and added until we arrived at the final set below in Table 3. The three annotators coded the channels in the sets a-h (as introduced above). In cases of disagreements, the coders again jointly discussed to arrive at a shared consensus.

**Table 3. Categorization of YouTube channel topics**.

| Abbreviation | Channel topics | Description and examples |
|---|---|---|
| STEM | science and engineering disciplines | This category covers channels that focus on academic content in the areas of science and engineering disciplines. It comprises formal and natural sciences such as physics, astronomy, chemistry, mathematics, statistics, geography, computer science, robotics. |
| BIO | biology and life sciences | This category covers channels that focus on academic content in the areas of biology and life sciences, including areas such as zoology (animals/species), botany, agriculture, genetics, immunology. |
| HUM_SOC | humanities and social sciences | This category covers channels that focus on academic content in the areas of social sciences and humanities, including for example economics, history, linguistics, |

| Abbreviation | Channel topics | Description and examples |
|---|---|---|
| | | theology, law, educational sciences, theology. It also includes academic perspectives on language and human behavior. |
| MULTI_SCI | multidisciplinary channels with professional science perspective | This category is used for channels that include academic content across several disciplines. The content is aiming at a primarily professional academic audience. |
| POPSCI | multidisciplinary channels with popular science perspective | This category is used for channels that present facts, interesting trivia, or selected academic findings in a lightweight style, addressing a general audience. Videos often have an entertainment element such as animated videos, reenacted experiments, narrators. |
| OPINION | personal opinion, commentaries and alternative news | This category is used for channels that are mainly distributing personal opinion video statements, or alternative andhyperpartisan news, or that discuss esoteric or religious world views and pseudoscience (from different perspectives, including debunking of pseudoscience). It also covers conspiracy beliefs expressed in some of the channels. |
| VLOG | personal vlogging / diaries | This category is used for channels that primarily report form a personal perspective on everyday life situations, often taking on the form of a video diary. |
| FIT | fitness, wellbeing, nutrition, beauty | This category is used for channels that cover topics around various aspects of body and soul, most notably fitness and dancing videos, tips for wellbeing, everything related to nutrition including cooking tips, beauty and fashion and other lifestyle tips. |
| HEALTH | health and medicine | This category includes videos on health and medicine related topics, including physiotherapy, pregnancy and birth, nursing, information about medical conditions, psychology, tips for medical support |
| POL | politics and activism | This category is used for all political topics, also including political activism (e.g. climate protest). |
| REL | religion | This category is used for channels with a focus on religious beliefs, also including discussions of, e.g., atheism or creationism. |
| CREATIVE | creativity | This category is used for channels that focus on creative activities, including music videos, cartoons or animations created by the channel owners. It may also feature creative activities such as crafting or knitting and may include "Do It Yourself" tips. |
| GAME | video games | This category is used for channels that focus on video games and gaming activities. |
| REVIEW | literature and media | This category is used for channels that focus on the discussion and reviews of literature (non-scientific publications, fiction) or movies, films, and tv series. |
| COACH | coaching and tutorials | This category is used for channels that aim to offer coaching, training, self-help instructions or tutorials. This can for |

| Abbreviation | Channel topics | Description and examples |
|---|---|---|
| | | example be tips for public speaking, psychology coachings, or language teaching classes. |
| OTHER/MISC | other topics or miscellaneous | This category is used for channels that focus on other topics not previously covered or on a variety of contents that cannot be clearly identified. It also includes channels that do not seem to include original content but mainly copy content from other YouTube channels or from mainstream media. |

# 4. Ethical Considerations

The main ethical concern of this paper is about the protection of YouTube users who have become the study subjects of this work. As in most research based on social media data this work has not been conducted on the basis of informed consent (Breuer et al., 2023). Lack of informed consent is almost impossible to avoid for large scale online platform data and for capturing unobtrusive online behavior for a specific research question. Therefore, other protection measures for platform users have to be explored in most cases.

This research is based entirely on data that is (or was at the time of data collection) publicly available information on either YouTube or other freely accessible Web resources. We only considered public/open YouTube channels. While for this research we used personal information (specifically names and information about educational and occupational background, such as academic degrees and affiliations with a university) in order to identify channel types, we only used this kind of information during the annotation process. The additional information was not stored as part of our dataset after the coders had agreed on a shared category for a channel. The annotation data however does still include YouTube channel names and channel descriptions and should either be anonymized or put under specific protection for future data sharing purposes[12]. Generally we can assume that most of the account holders in our dataset are aware of the publicness of their data on YouTube, and possess some medium to advanced level of using YouTube, since they are at least making use of video descriptions and including references in them. Still, one of the more vulnerable groups of people in our dataset turned out to be students, who have been assigned the task of producing and publishing a video as part of their degree programs and therefore can be assumed to be less voluntarily sharing their data on the platform. Therefore, preserving the anonymity of user accounts and removing all unnecessary information in this paper was our priority. For mentioning account names as examples in the analysis below, we focused on professional accounts (e.g., from publishers or universities) and highly popular accounts, and restricted us to paraphrasing when talking about the other accounts.

Another ethical consideration concerns the protection of the student assistant involved in the project as one of the coders for YouTube channel categories. Potentially the videos in our dataset could include disturbing contents of all possible kinds, that the coders would be exposed to without any content warnings. In our exemplary cases the most critical content types were opinionated videos that might include harmful speech towards minorities, but also explicit videos of medical procedures. We want to point out that studies following similar research designs should consider the possibility of facing more severe cases of problematic content. During our manual coding procedures we implemented practices to only open all YouTube links for inspection in a "private" browser window, not being logged into any Google accounts, for protecting coders private accounts.

---

[12] We are investigating options for sharing (anonymized) parts of our dataset with a research data archive and may provide raw data of individual coders assessments to the reviewers if needed.

# 5. Results

## 5.1. Videos and referenced publications

This subsection will focus on the videos themselves and the research publications mentioned within them. It will include an analysis of the types of publications that are commonly referenced, their fields of study, the citation impact of these publications, and any notable patterns or trends in how these publications are discussed or utilized in the videos.
To better understand how references to publications are distributed within YouTube (RQ1.1), we considered different dimensions. Specifically, we consider topical distributions of videos (via YouTube categories) and of cited publications (via Scopus subjects).

First, we consider the distribution across YouTube video categories: YouTube's API provides information on video categories assigned to the videos for organization within the platform. Figure 3 depicts the percentage of videos that fall into each video category[13], as well as the bootstrapped error for each sample. This data can help to differentiate which subgroup has significant differences compared to the full data or between subgroups. For example, even though in the full dataset there are around 20% of videos in YouTube's People&Blog category, the top viewed, commented, and liked videos subgroups are less likely to be in that category (with 1-4% videos). The videos with references to scientific publications are more likely to be from YouTube's Science&Technology, People&Blogs, or Education categories. At the same time, 50% of the most disliked videos are from the Education category (Figure 3), while in the full dataset, only 10% of videos are categorized as Education.

---

[13] In our plot, we showcase only the top 10 video categories. Additional categories like Music, Comedy, Gaming, Travel & Events, Autos & Vehicles, and Shows are present but constitute less than 1% of the videos, hence they are not displayed in the plot.

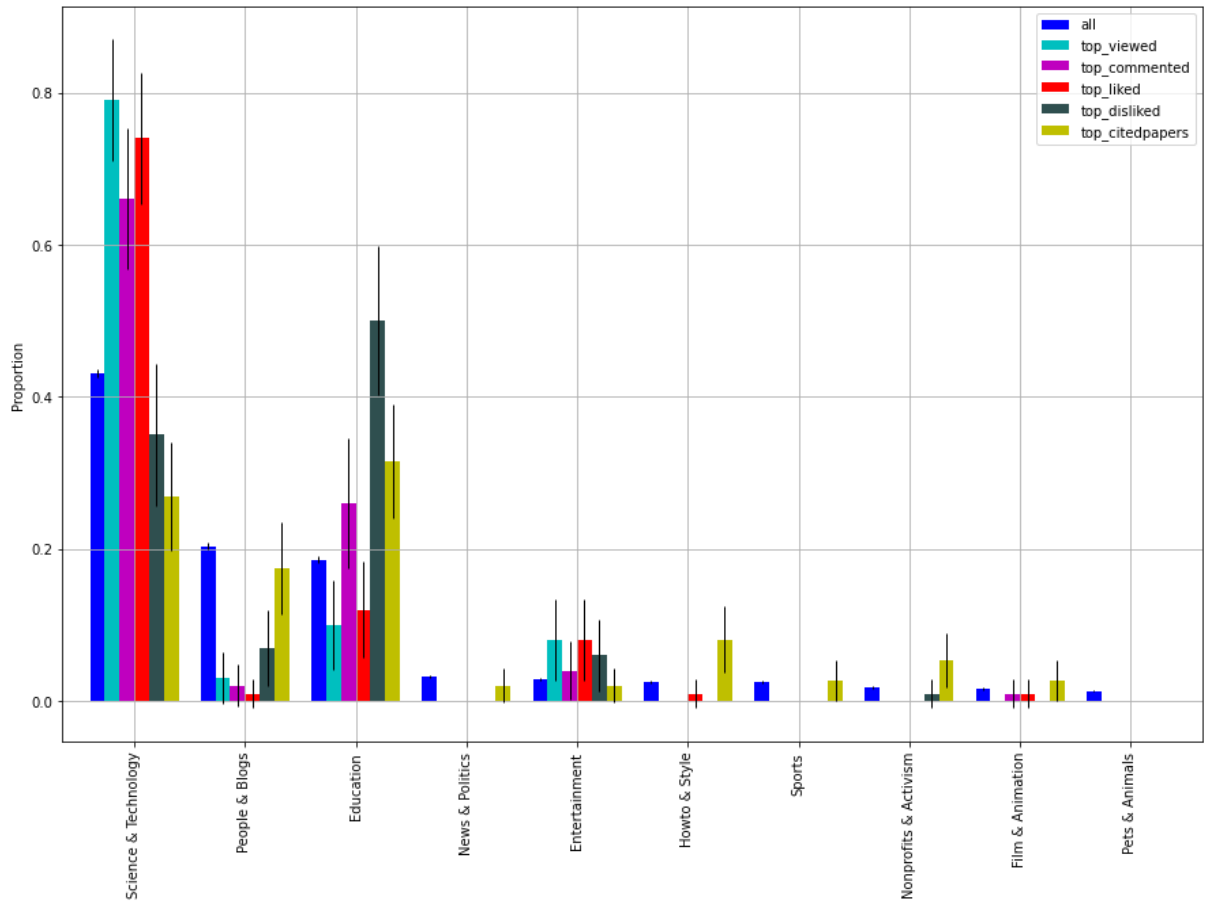

**Figure 3 - Distribution of videos in our dataset across the video categories assigned by YouTube**

For all the publications in our dataset (i.e., the publications that were referenced in at least one video description), we also considered Scopus subjects assigned to the respective publication. Figures 4 presents the frequencies of Scopus subjects in our full dataset and the subsets defined above. According to the obtained data (Figure 4), we will see more publications from Scopus' General subject category in the **top_citedpapers** subset than it would be expected in the full data. In our data, publications categorized under Scopus subjects such as Health Sciences, Medicine, Nursing, and Psychology are notably more prevalent in the **top_disliked** subset, i.e., subset of videos that received the most dislikes. Moreover, videos that reference publications in Pharmacology, Physics and Astronomy, and Physical Sciences are less likely to appear in the **top_disliked** subset than what random chance from full dataset would predict. **Top_liked**, **top_viewed**, and **top_commented** videos are more likely to have references from Social Sciences, Neuroscience, and Psychology than it would be expected (in the full dataset) by chance. At the same time, these three subsets (i.e., **top_liked**, **top_viewed**, and **top_commented**) are less likely to reference publications in Physical Sciences or Nursing.

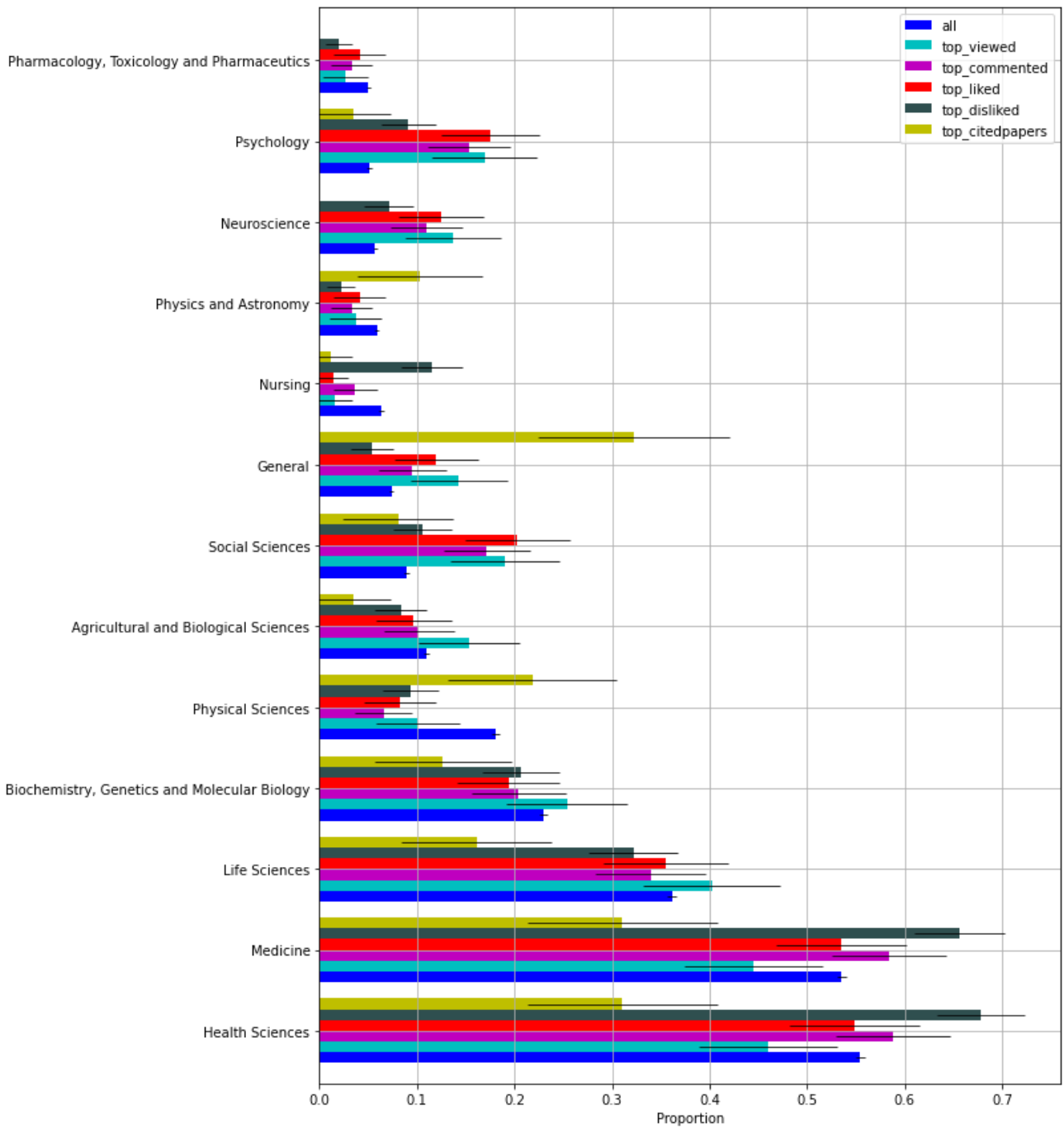

**Figure 4 - Scopus subjects of referenced publications in videos**

We also assessed whether there is a significant difference in citation impacts among various subsets. Based on the cumulative density plot, it appears that none of the subsets exhibit significant differences in the distributions of citation impacts within them (the only exception is the **top_citedpapers** which per design includes only videos with highly cited publications; Figure 5).

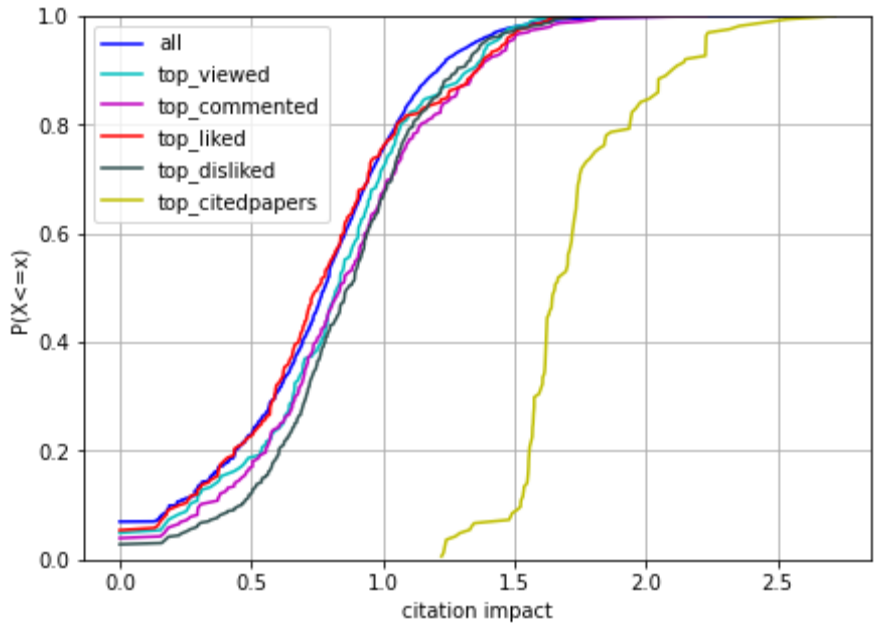

**Figure 5 - Cumulative density function representing the citation impact**

## 5.2 Characteristics of channel topics

This subsection will delve into the specific attributes of the channels that host these videos, based on our manual coding of YouTube channels that posted videos with references to publications in their descriptions. This analysis will include an examination of channel demographics, focusing on aspects such as their primary topics and the academic backgrounds of the channel hosts. We will also explore whether certain types of channels are more likely to reference research papers, and if so, what the nature of these publications tends to be.

The results of our coding, as depicted in Figure 6 under the chart labeled **random** show that approximately one-quarter of the channels are in the science and engineering (label: STEM) category (which includes several natural and applied sciences, like Physics, Engineering and Mathematics - but excludes, Biology and Life Sciences, and the Health domain which have their own categories). The Health category (label: HEALTH) appears as another main field in the random sample. Other frequent topics in the random set are from the area of Fitness and Wellbeing (label: FIT), with Opinion/Commentaries/Alternative News (label: OPINION) and Popular Science (label: POPSCI) channels also being notable. It is important to distinguish between channels focused on research in STEM fields and those that fall under the Popular Science (POPSCI) category: while the STEM channels focus on academic audiences and include content that resembles traditional academic output (such as recorded talks from conferences) POPSCI channels aim to simplify research topics for a broader audience and often present videos in a more popular style, using animations and narrators.

Comparing these findings with YouTube's automatic video categorization reveals a notable discrepancy: according to YouTube, around 42% of videos are from the Science and Technology category (as seen in Figure 3. However, YouTube's categorization does not differentiate between academia-focused videos and popular science or trivia videos addressing a general public within this category. Additionally, there may be individual research videos from science and engineering fields in the channels that we categorized as MULTI_SCI. This category encompasses channels that post research videos from various fields, not focusing on single disciplines. Examples include research institutions (e.g., Duke University) or publishers (e.g., WileyVideoAbstracts channel) that disseminate content across various fields such as humanities, social sciences, life sciences, medical science, and science and technology subjects. Publishers that focus on research within a single field (e.g., NewJournalofPhysics channel), would fall into our STEM channel category.

Another interesting category is the OPINION category, which includes a diverse range of channels. We've seen channels dedicated to esoteric topics, discussions on sexism, various conspiracy theories, and debates on subjectivity and objectivity or anti-scientific argumentations. In this category, there are channels that promote unscientific beliefs, as well as those that attempt to defend scientific perspectives and counteract fake or alternative beliefs (such as the flat Earth theory). Channels that present alternative or radical *political* beliefs were grouped into a combination of the two distinct categories for Opinions and Political topics ("OPINION; POL"). Most of these channels do not provide clear information about their creators and sometimes appear to be replicas of potentially banned channels or collections of videos about certain politicians curated by their fans. It should be noted that the random subset did not include any political channels from mainstream media.

When we reviewed all the publications referenced in our complete set of all videos, we found that between 53-55% of the publications are from the fields of Health and/or Medicine according to Scopus topical classification (Figure 4). These results are consistent with the recent findings presented in the study by Shaikh et al. (2023). However, channels specifically focused on Health and Medical research (label: HEALTH) account for only 11% of the random subset (Figure 6). While we cannot fully explain the size of this discrepancy, different factors may contribute to this. First it is possible that single videos in specific health related videos link to numerous publications, thus contributing to the high number of health and medicine related publications referenced from YouTube. On the other hand, publications in the medical domain may be referenced in videos from channels across different topical foci. In our case, the specific HEALTH category is used for channels that are explicitly dedicated to medical research and professional health perspectives. But the category is distinguished from topics around fitness (e.g., gym training videos, muscle increase), wellbeing (including relaxation and mental health tips) and nutrition (including tips and recipes for specific diets) which are collected in the as Fitness, Wellbeing and Nutrition (label: FIT). Examples of channels from the HEALTH category include:

- An international project providing education on research for birth workers (nurses, midwives, physicians).
- Channels belonging to research teams or divisions from biomedical research centers associated with certain universities and institutes.
- Channels of research journals from the Health and Medicine fields.
- Channels operated by medical students.

In contrast, the FIT category comprises:

- Channels promoting vegetarianism or veganism as a specific diet (by individuals or groups, regardless of their formal education).
- Video blogs on beauty and health maintenance.
- Channels by fitness trainers with instructions for building up muscles or losing body weight.
- Channels offering nutrition advice by nutrition coaches or (former) academics.

Channels from the FIT category are very likely contributing to the dominance of health and medicine related publications among all publications referenced from YouTube. For example, producers of fitness and nutrition related video content refer to medical findings that may support their recommendations and training advice. But medical literature may also be referenced from multi-topic channels, including popular science accounts (POPSCI).

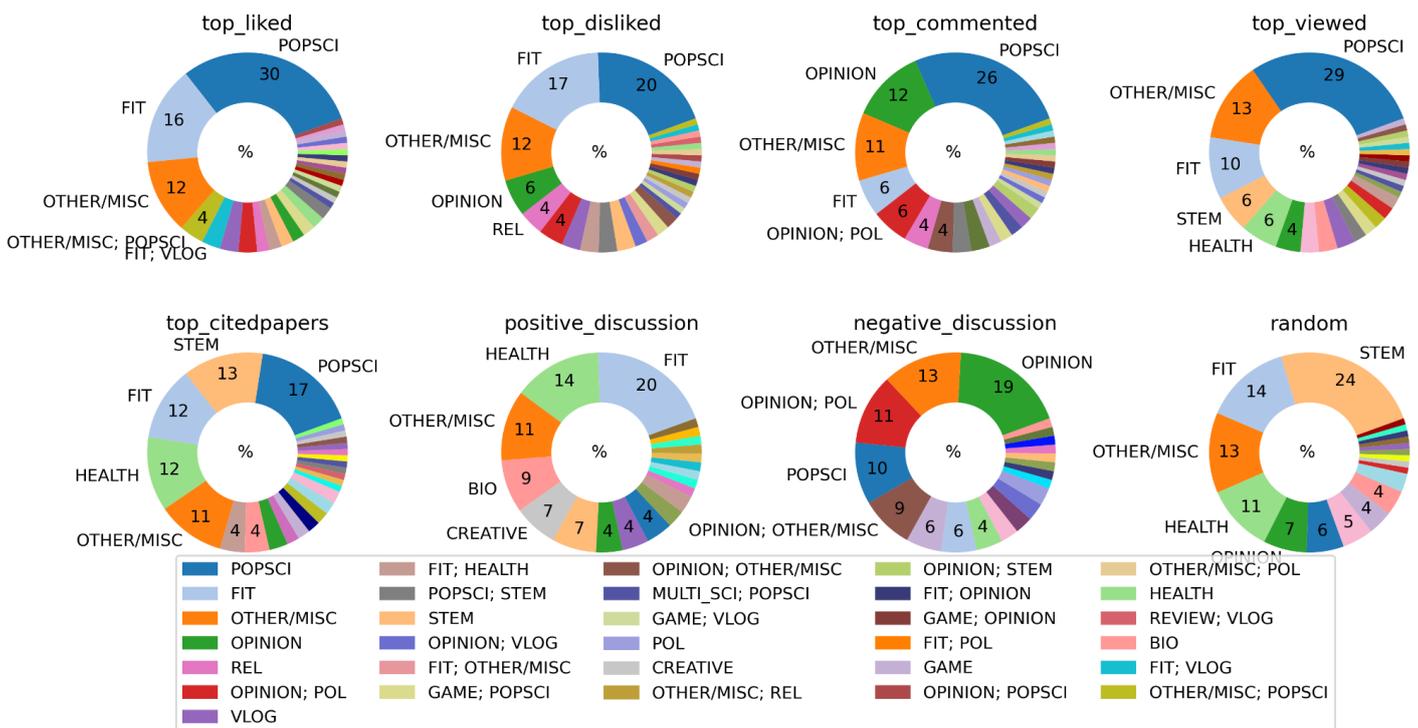

**Figure 6 - Video Topics of different channel subsets**

In addition to the random selection of channels, we can take a closer look at the subsets that encompass channels that have published particularly popular videos. When specifically looking into the **top_liked, top_disliked, top_commented, and top_viewed** subsets (Figure 6), these show a different composition of channel topics than the **random** subset. The most prominent channels now fall into the popular science domain (POPSCI: 20-30% across the different subsets), often followed by channels from the Fitness, Wellbeing and Nutrition domain (FIT: 6-17%). A group of unclassified channels (OTHER/MISC: 11-13%) should also be noted here briefly. In these subsets connected to popular videos the Other/Misc category includes channels that cover a variety of topics, such as trivia questions and everyday life facts (e.g., BuzzFeedVideo, LEMMiNO, Fatos Desconhecidos), and channels with a random assortment of videos lacking a clear topical theme (e.g., a mix of tech news, trivia, documentary films, and music clips). This latter type can also be classified as entertainment channels, which are typically large (in terms of the number of videos posted), have millions of followers and are managed by extensive teams. Channels that feature both trivia and popular science content are categorized as a combination of the two codes "OTHER/MISC; POPSCI" (1-4% in the popular videos).

Popular science channels are often well-represented and promoted on the internet, making information about team members and their biographies readily available. Based on our samples, popular science channels are predominantly run by large media groups or professional YouTube teams, with fewer of these channels being managed by professional academics affiliated with a research institute. But most of these channels include at least one individual with an academic degree in the field they discuss. For single-host channels in this category, we observed that the majority are science communicators with academic degrees (87% of them).

From the **top_commented** pie chart (Figure 6), we can see that the OPINION and OPINION; POL categories are overrepresented in this subset compared to other popular video categories. The group "OPINION" includes channels that for example: 1) rationalize various conspiracy theories (e.g., alien existence, 9/11 events, political conspiracies); 2) comment on or discuss social topics (e.g., criticize or defend feminism, LGBTI, gender or race issues), 3) attempt to debunk conspiracy theories or anti-vaccination movements. It should be noted that the code combination OPINION; POL is exclusively used for channels discussing (often polarized) political opinions, but without promoting political conspiracies (e.g., conspiracy theories about President Obama's birthplace fall under OPINION). In this study, we did not differentiate between defenders and deniers of scientific approaches. However, the majority of videos in the top_commented category (unlike the random or negative_discussion subgroup) appear to be from defenders of the scientific approach, aiming to counteract conspiracy beliefs. Uncovering the exact relation between conspiracy beliefs and references to academic publications on YouTube would be a potential topic for future investigations.

When we look into the **top_citedpapers** subset, we take on the perspective of the cited publication. Here it is noteworthy that videos mentioning publications with high citation impact show a similar consistency of topics compared to the random subcategory. However, there are notable differences: 1) POPSCI channels are the most prominent group (17%) in the top_citedpapers subgroup, 2) OPINION channels are almost nonexistent in the top_citedpapers, and 3) STEM channels represent only 13% of the top_citedpapers subgroup, as opposed to 24% in the random category. Thus, we can hypothesize that highly cited publications are more likely ($X^2$ (1, N = 100) = 21.5, p < .00001) to be referenced by Popular Science channels and less likely ($X^2$ (1, N = 100) = 6.3, p < .00001) to be referenced in STEM content than would be expected by chance.

The distribution of topics within the **negative_discussion** and **positive_discussion** channel subgroups deviates from the patterns observed in all popular and random subgroups. It is noteworthy, however, that the "OTHER/MISC" topic remains consistent across these groups, maintaining the same relative frequency (11-13%). We observed that the top_commented subgroup contains a significant number of channels within the OPINION category, and this trend is even more pronounced in the negative_discussion subgroup. In this latter subgroup, the most frequent topics are OPINION (19%), "OPINION; POL" (11%), "OPINION; OTHER/MISC" (9%), POL (6%), and GAME (6%). Popular Science channels, while present in the negative_discussion subgroup at 10%, are not as prominent as they are in the top_commented subgroup, where they account for 26%.

In contrast, channels within the **positive_discussion** subgroup tend to focus more on Fitness and Nutrition (FIT), Health and Medicine (HEALTH), Biology research fields (BIO), and Creative activities (CREATIVE). The prevalence of the CREATIVE category among the videos with specifically positive comments is understandable, as audiences often express appreciation for creativity, whether in DIY tips and tricks or in showcasing talents like music videos or animation creation. However, it is particularly interesting to note the positive reception of topics related to biology and the human body, as evidenced by the rewarding comments they receive. In Zagovora & Weller (2018) we presented initial examples of qualitative analyses from the comments.

## 5.3 Characteristics of channels owners

After focusing on the channel topics in the previous section, we now look at the persons behind the channels. This enables us to get an insight into whether engagement with scientific publications on YouTube goes beyond core academic actors. As a general observation we may already conclude from our data that sharing of references in video descriptions is happening also by a general YouTube community that is not affiliated with academic institutions and also does not display any connections to having an academic degree or background.

Even so academics, research institutions, and publishers have a notable presence in the channel dataset, as evidenced by their respective shares of 25%, 10%, and 9% in the **random** subset (Figure 7), they do not achieve the same level of popularity as other types of channels. The groups comprising the most popular channels (**top_liked**, **disliked**, **commented**, and **viewed**) are predominantly occupied by professional YouTubers, both with and without academic backgrounds, as well as commercial media.

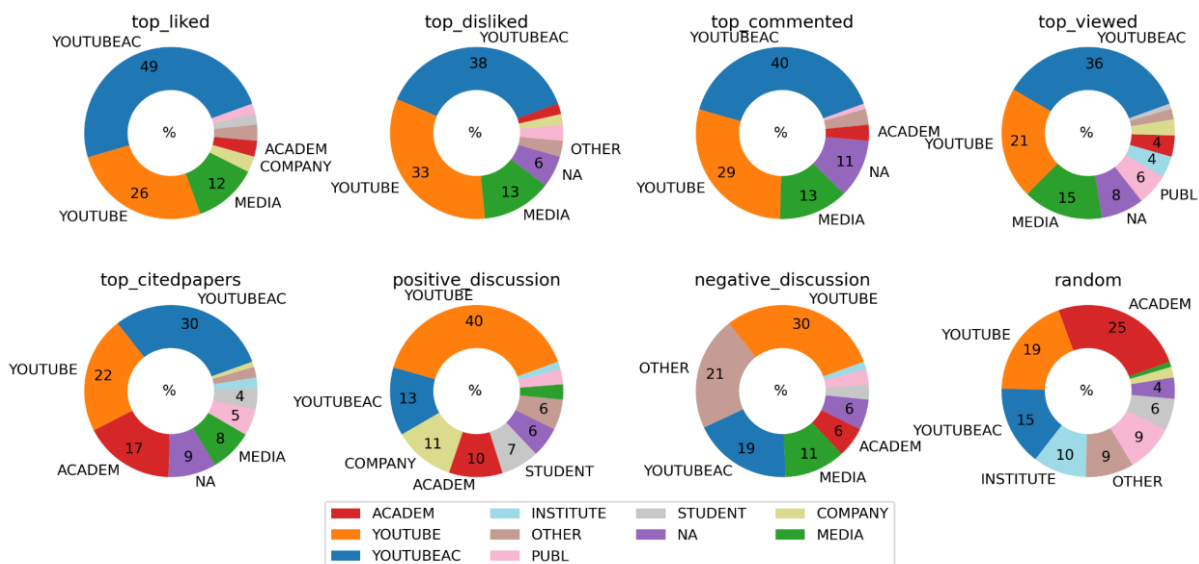

**Figure 7 - Channel types based on background of their owners**

This disparity in popularity may be attributed to the substantial effort and resources, including both time and money, required for effective promotion on YouTube. To achieve trending status, individual academics would likely need to either hire professionals for social media management and video production or shift their focus entirely to becoming close to full-time YouTuber content producers. In the latter scenario, they would then likely appear in our dataset in the YOUTUBEAC category.

It might seem surprising to see that even within the **top_citedpapers** subgroup, academia-related channels are underrepresented compared to what we would expect based on a random distribution. We had initially expected a higher prevalence of academic channels in this subgroup. Consequently, this led us to conduct a further investigation into the citation impact of publications referenced by academics, institutions, and publishers, which is detailed in Subsection 5.4.

In the **negative_discussion** subgroup, we observed a disproportionately high number of OTHER channels and YouTubers without academic backgrounds. The channels classified as OTHER often do not disclose their identities but may work with aliases, and in many instances, it's unclear whether the content they publish is original. They do however include some sort of consistent account appearance, e.g. a channel description or motivation. In contrast channels completely lacking a description, name, appearance of a person behind them in some of the videos, and almost no context were classified as NA in our study. Channels in the OTHER category may also focus on specific topics, which in our subset include, amongst others: alternative news media, channels featuring collections of documentary style videos about national histories, and single cases of discussions that include problematic and harmful online content such as racism.

Interestingly, within this diverse group of OTHER channels, we also identified three channels that posted supplementary material (in video form) to actual research publications. While it is likely that these may have been the authors of the respective published papers, no individual or group could be definitively identified as the owner(s) of the YouTube channels behind these three videos. Their appearance of these videos in the **negative_discussion** subset however can be explained: the videos as supplementary material for published papers included very explicit material, e.g. from medical procedures or biological experiments, that caused reactions of disgust in the comment sections.

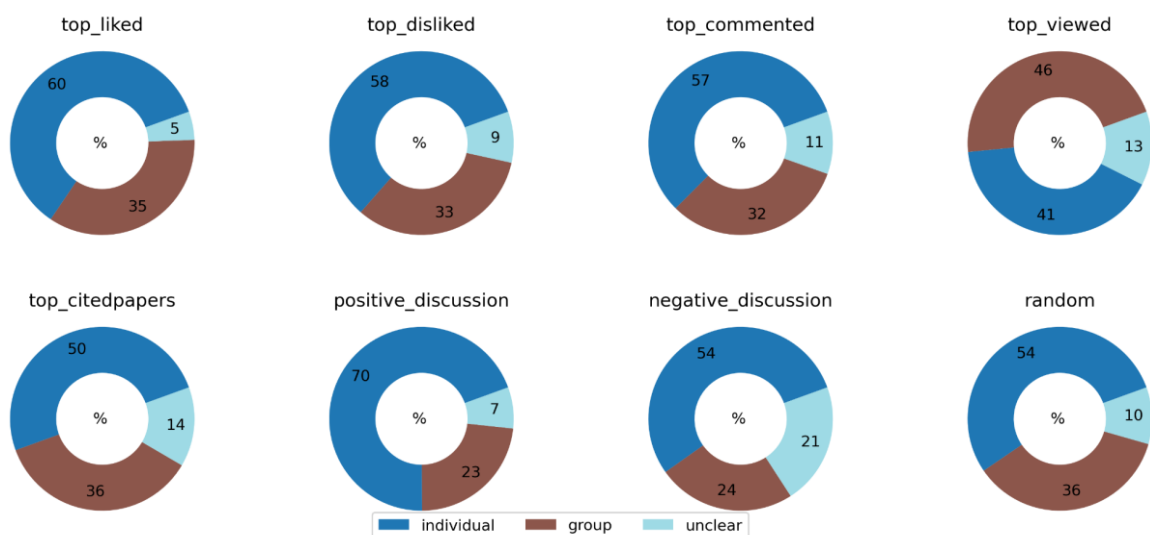

**Figure 8 - Structure of channel owners**

Finally, for our question of who is behind the channels that post videos with links to publications, we looked into whether the channels were operated by individuals or groups. In all subsets (Figure 8), with the exception of the top_viewed category, the majority of channels are led by individuals, ranging from 50 to 70%. As previously noted in our channel category analysis, the negative_discussion subgroup distinctively includes more channels in which the owners choose not to reveal their identities. For approximately 21% of the channels in this subsets it was not even possible to detect if the channel was operated by an individuals or a group of people.

## 5.4 Channels by academic actors

In this subsection, we closely examine the content published on YouTube by scholarly actors, including academics (label: ACADEM), research institutions (label: INSTITUTE), and publishers (label: PUBL). For these groups we investigated more closely to better understand the following two aspects: first, the range of topics covered by their channels, second the citation impact of the publications they reference. For this part of the study, we exclusively considered our random subset of channels as introduced above, and focused on the academic actors there (44 channels: 25 ACADEM, 10 INSTITUTE, 9 PUBL, compare Figure 7). This approach was chosen to avoid potential biases that might arise from the nature of subgroup creation, ensuring that the inferences drawn about these channels are not influenced by the specific characteristics of any one subgroup.

As evidenced in Figure 9, academic actors display a distinct distribution of topics compared to random channels, as shown in Figure 6. Notably, researchers (label: ACADEM) that are setting up channels connected to their professional career appear to use these accounts almost exclusively for professional academic topics. In the topical distribution in Figure 9 we no longer encounter any channels focusing on opinions and debates (OPINION), or on fitness and lifestyle (FIT). We still see some contributions to conveying scientific findings to a general public with 8% of channels featuring POPSCI content. But overall, the focus clearly is on research content. Within this clear focus on academic topics, we also observe a very strong preference for natural science, technology, and health topics, with 60% in the STEM category, 16% in BIO, and 8% in HUM_SOC, 4% in HEALTH. While our approach does not enable us to explain this skew towards natural science disciplines, it could be interesting as a starting point for future research. It may well be an indicator that different disciplines develop different practices for using YouTube as an online platform, but we cannot state from our data whether this is potentially because some disciplines act as early adopters of this platform, whether science communication plays a different role for them, or whether they apply different citation practices in online environments (and therefore are more or less likely to be in our dataset that roots in the existence of a reference to a publication in a video description).

For our channels operated by research institutes (label: INSTITUTE), we see a larger share of the HEALTH category (30%) compared to the individual researchers (4% in HEALTH). Publishers' channels (PUBL) present an even stronger focus on HEALTH topics (56%) on the one hand, while on the other hand being mainly multidisciplinary (33% in MULTI_SCI, thus not focusing on a specific academic field). The MULTI_SCI; OTHER/MISC category (10% for INSTITUTE channels, Figure 9) encompasses research institutes that feature a diverse range of content on their channels. This includes not only academic research but also aspects of student life, campus events, and other related topics.

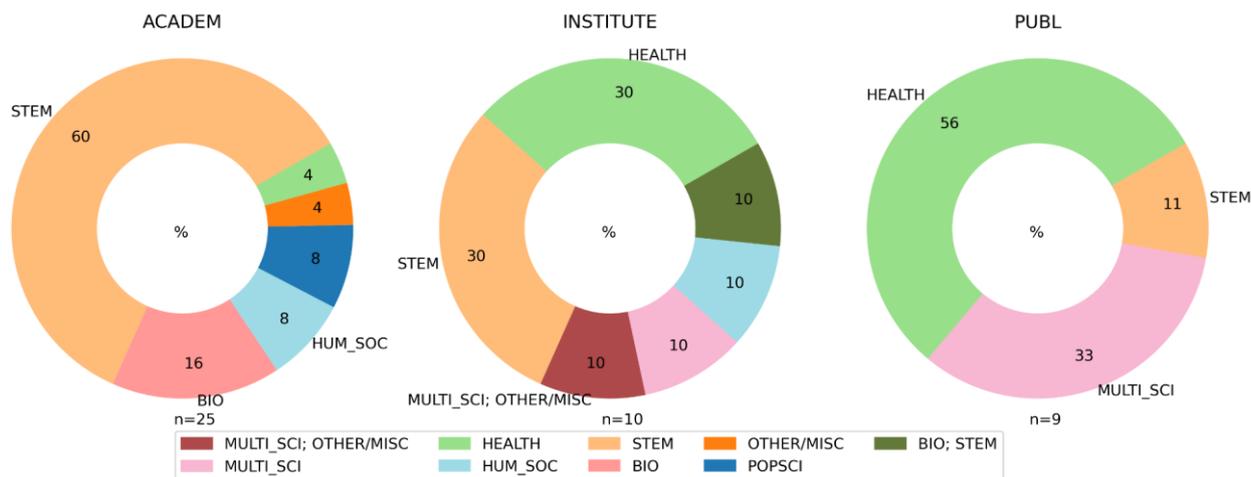

**Figure 9 - Channel topics of academic actors (academics, research institutions and publishers)**

Two quarters of the channels in the ACADEM category are managed by individual researchers, while around 30% are owned by small teams of academics. By design, larger entities, such as research divisions, are categorized under the INSTITUTE category. For publishers' channels, they are all coded as group-managed entities. This classification is applied regardless of whether a single individual physically manages the channel, which is often not verifiable without direct inquiry. This coding approach acknowledges that these channels represent the collective efforts of the larger group involved in journal publishing activities.

As we have seen, academic actors on YouTube appear to focus on academic topics in their videos. But can we rely on academics on YouTube for identifying highly cited publications? For instance, Zahedi et al. (2015) demonstrated that readership scores from Mendeley can be effective in pinpointing such impactful, highly cited publications. As a first test to this hypothetical connection in the context of YouTube channels, we calculated the mean citation impact for every channel in our random subgroup. Figure 10 presents these numbers, grouping channels by category. As observed from the plot, channels categorized as INSTITUTE, OTHER, YOUTUBE, and YOUTUBEAC demonstrate a higher average citation impact in the publications they reference in their videos descriptions compared to those labeled as ACADEMICS. However, it is apparent from the plot that these differences are not statistically significant. Therefore, a much larger random subset is needed to robustly test whether non-academic YouTubers or Institutes can be employed as effective filters for identifying highly cited research. Nevertheless, the plot also indicates that both academics' and publishers' channels tend to reference publications across the full spectrum, suggesting they cannot be reliably used for the purpose of predicting impactful publications. One plausible explanation is that researchers and publishers use YouTube to articulate and disseminate their own research findings, without pre-selecting those that have already achieved a certain level of citation impact. Another possibility is that researchers often publish very recent findings on YouTube, which have not yet had the opportunity to accumulate citations.

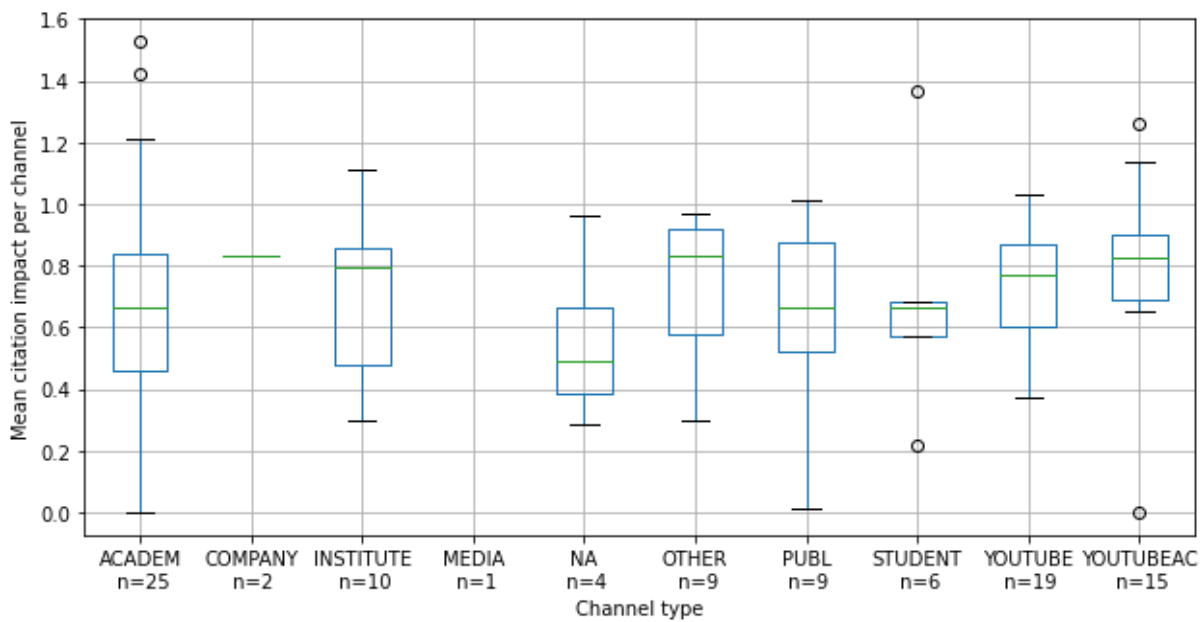

**Figure 10 - Mean citation impact of referenced publications by channel categories**

# 6. Discussion & Conclusions

**YouTube categories, Scopus subjects, and channel topics.**
The analysis reveals a significant presence of health and medicine-related publications in YouTube content, accounting for 53-55% of all referenced publications, despite health-focused channels comprising only 11% of a random subset. This discrepancy suggests a broad interest in and utilization of health and medical content across various types of YouTube channels, not just those specifically dedicated to health and medical research. Channels under the Fitness, Wellbeing, and Nutrition category, as well as those covering a range of topics including popular science, contribute to this trend by frequently referencing medical literature, indicating a widespread integration of health and medical information into diverse video content on YouTube. This reflects the platform's role in disseminating health and medical knowledge beyond traditional academic and professional circles. This trend can inform bibliometrics research by highlighting the cross-disciplinary impact of research publications, an aspect often overlooked in traditional citation analysis.

From the channel perspective, our findings highlight a diverse landscape of YouTube channels, with a significant presence of science and engineering and health-related content, alongside popular science, fitness & wellbeing, and opinion & commentary channels. This diversity underscores YouTube's role as a multifaceted platform for both academic and popular dissemination of knowledge. A notable discrepancy exists between YouTube's broad Science and Technology categorization and the more nuanced distinctions in the study, revealing the platform's limited ability to differentiate between academic-focused and general audience content. Thus, we would not recommend to use YouTube video categories in a similar way as we would consider fields of studies in bibliometrics. Rather additional categorization of content is needed to differentiate between scientific videos and educational. Additionally, the opinion & commentary channels' varied content, including channels promoting both scientific and unscientific beliefs, reflects the platform's complex role in shaping public discourse on scientific and political topics.

Popular science content, along with fitness and wellbeing topics, garners significant engagement and popularity on YouTube. These contents on YouTube highlight the platform's role as a significant medium for education and lifestyle enhancement. It also underscores the public appetite for content that is both informative and practical in these domains.

The opinion & commentary and opinion & commentary in politics channels are significantly represented in the top_commented subgroup and even more so in the negative_discussion subgroup, suggesting that opinionated content, particularly involving political or controversial topics, drives higher engagement, especially in negative contexts. In contrast, fitness, health, and biology-related content, along with creative activities, receive a positive reception, reflecting audience preferences for content related to well-being and human interest topics.

There is a clear connection between the nature of the YouTube channel and the type of academic publications it references, with popular science channels more likely to reference high-impact research. This suggests that popular science channels play a significant role in disseminating high-impact academic research to a broader audience.

**People behind channels.**
Moreover, popular science channels are predominantly managed by large media groups or professional teams, with a notable presence of science communicators holding academic degrees. This suggests a professional and structured approach to science communication in popular videos.

Despite the presence of academic (25%), research institution (10%), and publisher (9%) channels in the dataset, the most popular YouTube channels (top liked, disliked, commented, and viewed) are predominantly occupied by professional YouTubers and commercial media. This suggests that non-academic content creators with or without academic backgrounds are more successful in gaining popularity on YouTube. The effort required for effective promotion on YouTube might be a significant barrier for individual academics and research institutions. The dominance of professional YouTubers and commercial media in the most popular channels, despite the presence of academic, research institution, and publisher channels, highlights a gap in how academic content is promoted and received on social media. This could lead to further research on strategies for academic entities to effectively engage audiences on platforms like YouTube.

Across most subsets, the majority of channels are operated by individuals, with a significant proportion in the group of channels with negative discussions being anonymous or unclear.

**Academic actors.**
Scholarly actors on YouTube have a distinct topical focus compared to the general channel population, with a strong inclination towards academic and research content. The focus on natural science disciplines among academic channels suggests potential discipline-specific patterns in the adoption and use of YouTube for science communication. At the same time, Research institutes and publishers show a larger share of health-related content compared to individual researchers.

The unique focus of academic actors on YouTube, particularly their inclination towards natural sciences and health-related content, reveals discipline-specific patterns in science communication. This could be a vital consideration for altmetrics, suggesting a need to explore discipline-specific altmetric indicators.

Academic actors on YouTube display a focus on academic content, but their effectiveness in identifying highly cited research is uncertain due to the broad range of publications they reference and the lack of significant differences in citation impact across different channel categories. Further research with a larger sample size is necessary to determine the potential of professional YouTube channels (versus academic actors) as a tool for identifying impactful research publications.

## 7. Author contributions

Conceptualization: Olga Zagovora, Katrin Weller; Methodology: Olga Zagovora, Katrin Weller; Formal analysis and investigation: Olga Zagovora, Katrin Weller; Writing - original draft preparation: Olga Zagovora, Katrin Weller; Funding acquisition: Katrin Weller; Resources: Olga Zagovora, Katrin Weller; Supervision: Katrin Weller.

## 8. Statements and Declarations

The authors have no relevant financial or non-financial interests to disclose.

## 9. Acknowledgments

The research leading to these results received funding from DFG - German Research Foundation under Grant Agreement No 314727790. The authors wish to thank Altmetric.com for providing this study's data free of charge for research purposes. We thank our student assistant Christine Gujo for their help during the coding process.

This is a preprint version.